# RSAM: An Enhanced Architecture for achieving Web Services Reliability in Mobile Cloud Computing

**Amr S.Abdelfattah**
Computer Science Department,
Faculty of Computers & Information Sciences, Ain-Shams University
amr.elsayed@cis.asu.edu.eg

**Tamer Abdelkader**
Information Systems Department
Faculty of Computers & Information Sciences, Ain-Shams University
tammabde@cis.asu.edu.eg

**EI-Sayed M. EI-Horbaty**
Computer Science Department
Faculty of Computers & Information Sciences, Ain Shams University
Shorbaty@cis.asu.edu.eg

**Abstract—** The evolution of the mobile landscape is coupled with the ubiquitous nature of the Internet with its intermittent wireless connectivity and the web services. Achieving the web service reliability results in low communication overhead and retrieving the appropriate response. The Middleware Approach (MA) is highly tended to achieve the web service reliability. This paper proposes a Reliable Service Architecture using Middleware (RSAM) that achieves the reliable web services consumption. The enhanced architecture focuses on ensuring and tracking the request execution under the communication limitations and service temporal unavailability. It considers the most measurement factors including: request size, response size, and consuming time. We conducted experiments to compare the enhanced architecture with the traditional one that has no additional overhead. In these experiments, we covered several cases to prove the achievement of reliability. Results also show that the request size was found to be constant, the response size is identical to the traditional architecture, and the increase in the consuming time was less than 5% of the transaction time with the different response sizes.

## I. Introduction

The recent explosion of the cloud computing is facilitating the deployment of web services, such that the web services act as self-contained components, which are published, located and invoked over the Web. The web services are the perfect way to provide a standard platform and operating system independent mechanism. This triggered the widespread usage of mobile applications over the Web.

The consumption of web services through mobile client prefers the choice of Representational State Transfer (REST) service type. This is because REST architecture is fundamentally client-server architecture, and is designed to use a stateless communication protocol, typically HTTP [1]. In REST architecture, clients and servers exchange representations of resources using a standardized interface and protocol. These characteristics encourage REST applications to be simple and lightweight. Therefore, regarding the scope of reliability, RESTFUL services overcome Simple Object Access Protocol (SOAP) services limitations as mentioned in [2]-[3] and achieve better results, especially in mobile communications [4]-[6]. REST services use HTTP request and response, which means that a mobile device connected with the internet can access the service without additional overhead, unlike SOAP web services [7]. According to [8], combining RESTFUL design with other technologies such as caching provides good system scalability.

Cloud Computing uses web services for connections. Most of these cloud oriented services and data are deployed as web services that are network-oriented applications [9]. The synchronization between a mobile client and a web service is achieved through initiating a conversation in a request response pattern.

Cloud computing improves scalability and consistency of services and data, and facilitates the deployment of mobile applications [10]. Therefore, the consumption of these services and pool of data and information is affecting smart phones to gradually become the effective client platform to consume.

Mobile Cloud Computing (MCC) is the combination of cloud computing, mobile computing and wireless networking to bring rich computational resources to mobile users, network operators, as well as cloud computing providers [11].

The uncertainty of mobile connectivity results in less satisfaction for the mobile user. In addition the network bandwidth limitation and unreliable wireless communication are decreasing the overall support for web service consumption on mobile devices [2]. The limitations are listed in the following points:

- Client Connection loss: The mobile clients have intermittent connection because of their mobility. They can be momentarily removed from the connected network and later join the available network [7].

- Service Connection Loss: The Cloud/Server may lose the connection and their deployed web services become unreachable from clients.

- Server Error: The Cloud services may temporarily suffer from unexpected error, which may be produced because of high load requests, or system environment issues. This will overload the mobile client to reconstruct and resend the request later.

- Bandwidth limitations: Cellular networks have a very limited bandwidth that may cause slow service consumption or request timeout response.

- Reliability: The public Internet is unreliable, such that if a client calls a web service and there is no response within the timeout period, the client aren't sure whether the web service request was successfully received, was lost in the internet before reaching the server, or was partially processed. In the case the application retries the operation and resends the request, it may be duplicated or cause an error, such as two orders entered or two credit card charges.

- Longer Transaction time: Web service consumption will be slower because of the HTTP overhead, the XML overhead, and the network overhead to a remote server. Therefore the differences in performance need to be factored into the application architecture to prevent unexpectedly poor performance due to the latency of the web services consumption failures.

The reliable web services consumption through mobile cloud computing is achieved by ensuring the appropriate response comes from the request execution for the service deployed over cloud computing. This execution can be under the intermittent connectivity, services unavailability, and the other unexpected conditions that will be covered in later sections.

In a highly distributed system where web services are scattered across multiple platforms, three system guarantees are required: Consistency of the data, Availability of the system/data, and Partition tolerance to fault. However, the "CAP theorem" [12] – [13] states that at most only two of the three can be guaranteed simultaneously. In distributed mobile systems where the mobile node is employed as the client platform of the web services, partition tolerance is a given because of the intermittent connectivity losses. This means we are forced to choose between Availability and Consistency.

The proposed approach is concerned with optimizing the request behavior rather than the request structure to achieve the REST web services reliability through mobile cloud computing. This approach defines the middleware as two components: one in the mobile client side and the other in gateway between the mobile client and the cloud service side. The integration between these both components achieves the request and response state awareness from the mobile and cloud service sides. These integration is used to make both the mobile client and the middleware component are aware with the state of each request with its corresponding response from the cloud service, such that the connection with the cloud service is managed through these middleware components. The caching technique is embedded to reduce the request timeout problem and reduce the overall requests to the cloud services. Though this approach introduces a new mobile application experience technique that either ensures the request execution or informs the user with the request state to take the suitable action regarding the failure states.

The rest of this paper is organized as follows: Section II contains an overview of the related work. In section III, the middleware approach for achieving the reliability is shown. The proposed RSAM architecture and its components detail is explained in section IV. The RSAM protocol and reliable consumption analysis is figured in section V. The environment setup is shown in section VI. Section VII discusses and analyzes the conducted results. The conclusion and future work are given in section VIII.

## II. Related Work

The middleware approach is applied in more than context with different purpose, such that Table 1 shows the usage summary of the middleware with different web services types and its utilization of caching technique. The Caching technique is the main technique that middleware rely on it. The caching technique is created in the middleware and the mobile device, the middleware caching is responding with the last saved results in case of cloud service unavailability, and the cache in the mobile device is used for showing the last results in case of internet connection unavailability. A notification mode was saved in the middleware to notify the mobile device once it becomes online again.

Table 1: The Comparison summary of the usage of the middleware approach

| Reference | Service Type | | Cache | |
|---|---|---|---|---|
| | Rest | SOAP | Mobile | Server |
| Consuming Web Services on Mobile Platforms [2] | No | Yes | No | No |
| Enhance the Interaction Between Mobile Users and Web Services using Cloud Computing [7] | Yes | Yes | Yes | Yes |
| Reliable consumption of web services in a mobile-cloud ecosystem using REST [10] | Yes | No | Yes | Yes |
| The proposed Architecture | Yes | No | Yes | Yes |

The middleware is tailored regarding to the required case in each of [2], [7], and [10]. In [2], the author is concerned with overcoming the heavy SOAP requests by converting them to lightweight requests. The request flows as following: the mobile constructs the intended request with binary protocol or REST then sends it to the middleware that reconverts it to its original SOAP request and send it to the intended service. Regarding the response flow, the service response result is sent to the middleware to convert it to the equivalent lightweight format like: JSON then respond with it to the mobile device. The failure is overcame by creating a module to retry the request invocation based on the saved state when both cloud service and mobile device are online.

In [7], the author focused on two main points. The first is converting the heavy request to lightweight one, such that in the case of SOAP service, the middleware converts the SOAP to REST request then converts the heavy XML response result to JSON format. The second is creating a caching module in both middleware and mobile device, and consider some requests to be connected directly to the cloud service to optimize the request and the response time.

In [10], it is applied for achieving the Synchronization concept between cloud data and mobile device data. They followed the CAP Theorem [12]-[13] by choosing the Availability and Partition tolerance as main points with slight consistency (Session Consistency), and it consider the caching technique in both sides.

Each of the above references discusses a different case using the common middleware approach. Through the proposed approach uses the middleware approach as the above references, but it discusses a new and different case. The proposed approach updates the middleware definition as two integrated components: the first is included the mobile client side and the other is in the gateway between the mobile client and the cloud service side. This approach considers the REST web services reliability through mobile cloud computing by optimizing the behavior of the request rather than the request structure in order to be aware of this request state under the connectivity limitations.

### III. Reliable Web service Using Middleware Approach

The middleware component acts as a gateway that communicates lightly with the client while ensuring the responsibility of retrieving the response from the web service.

The proposed communication architecture introduces a gateway between the mobile client and the web service that takes all the burden of the heavy load communication with the service. The mobile client will instead have to sustain a lightweight and simple client-server communication over a fast binary protocol [10]. The future middleware solution for mobile clients mostly focuses on application and content adaptation.

The four main middleware fundamental requirements are Heterogeneity, Network communication, reliability, and coordination. Scalability can be achieved with distributed middleware. Context can help middleware adapt to the heterogeneous environment. The objective of the middleware is to improve the interaction between mobile clients and web services. The cloud platforms support the middleware objective, such that it used to improve the scalability and reliability of the middleware.

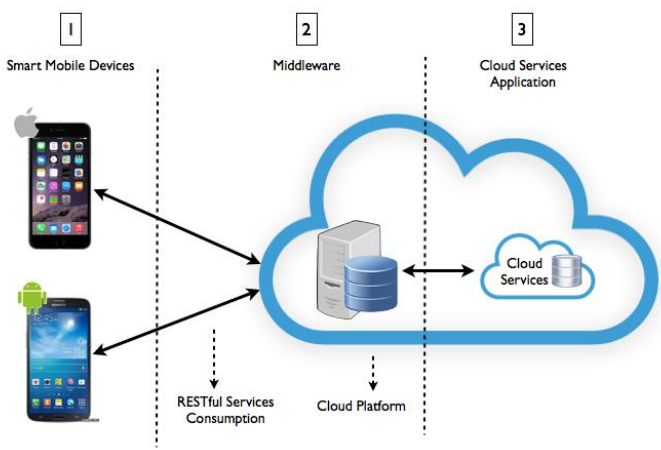

Figure 1: Middleware Architecture

#### 1) Middleware Architecture

The Middleware architecture, as shown in Figure 1, consists of the following three components:

1. **Smart Mobile Devices**: Mobile devices that have Internet access to be able to consume the web services.
2. **Middleware**: A web application that contains web services consumption, data format and handlers libraries.
3. **Cloud Services Application**: A web application that contains web services connected to a database to store the needed data.

#### 2) Middleware Advantages

The Middleware advantages can be summarized as following:

- Small bandwidth usage for limited GPRS data communication because of the light communication with the mobile client
- Little chance for failure in unreliable wireless networks.
- Takes the load of retrieving the response from the Web service.
- Advanced security features can be explored as the middleware acts on behalf of the mobile client.
- This architecture brings more opportunities towards ensuring a more reliable communication with the Web service such as:
    o The middleware will most likely run on dedicated hardware that will justify the search of solutions to ensure some kind of state of the communication with a Web service.
    o Retry mechanisms can be explored in case of connection failures.
    o Some communication states can ensure the transparency to the mobile client, such as in the case of communication failures (mobile device to middleware or middleware to Web service), the middleware can retain the state of the overall communication and retry to continue when all parties come back online.

#### 3) Middleware Limitations

The Middleware limitations are:

- The increasing of the overall duration of a request to the Web service, but this can be enhanced by deploying the middleware and the actual Web service in the same network (Via LAN).
- The heavy HTTP communication from middleware to Web service will bypass the network firewall thus adding an extra boost in the communication.
- The system might consume some time to process the data format on the middleware, but it may be lost because two communications lines have to be prior established and maintained.

### IV. Proposed Reliable Service Architecture using Middleware (RSAM)

The enhanced middleware architecture focuses on the integration between the mobile client and service layer, so the architecture defined as client middleware component and service middleware component. This integration increases the system awareness for each request state to be notified to the user appropriately.

In this architecture the mobile consumer component has the permission to consume a cloud service directly without passing through the middleware, which creates the flexibility to customize system communications.

The Enhanced architecture components and their communications shown in Figure 2, contains two main components:

1. **Cloud Service Consumer** is a mobile client middleware component responsible for:
   - Constructing the request with its appropriate attributes to be ready for sending.
   - Handling the request communication cycle either to the cloud service directly or through the service middleware component.
   - Receiving the response and notifying the client with the appropriate state.

2. **Middleware Service Component** is a cloud service middleware component responsible for:
   - Receiving the request from the client service consumer, and constructing the appropriate response regarding the sent request attributes.
   - Communicating with the required Cloud Service.
   - Caching the request and response states to be tracked for later usage.

A detailed description of the architecture is in the following subsections.

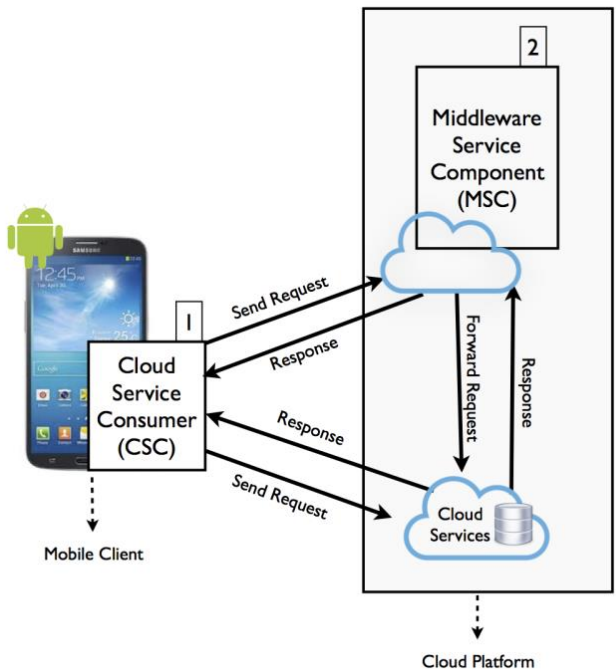

Figure 2: RSAM Architecture

**IV.1 Cloud Service Consumer (CSC)**

The CSC is the Client Middleware Component responsible for handling the request communication cycle starting from the service call till the response notification.

There are attributes that affect the consumer and support it to track and handle the consumption process as a separate component, as shown in the following:

- **Service Descriptor** is the module that contains the routing mechanism regarding the base cloud service URL and the middleware component URL, and has the description of each service in the system. This descriptor consists of:
  - **Basic Attributes** that describe the services, such as the service name, request type, parameter names, parameter types, and expected response type with its appropriate parser.
  - **Sematic Attributes** that affect the behavior of the execution, such as:
    - The forced attributes that force the MSC to use the succeeded cached results, or forward the request to the cloud service each time.
    - The direct attribute that controls the request route to the middleware or to the cloud service directly.

- **Request Client Id** is the unique identifier over all requests in the system, it consists of two main parts,

$$Uid = S + b; S = d + t + r$$

where, **Uid** is the unique identifier to be used as the request client id.

**S** : The device unique structural part.

**b** : The part that is unique regarding the behavior of the request itself.

**d** : The device manufacturer identifier.

**t** : The time of the sending the request.

**r** : The requested service url.

**+** : Concatenation operation.

**,** Such that the above equation produces the "Uid" Unique Request Client identifier to be ready send it to the middleware.

  - **Basic Part** is the auto generated unique key that consists of subparts that ensure its distinct property, such as mobile device manufacturer number, the request timestamp, and the requested service name.
  - **Additional Attributes Part** is the other part that includes labeled attributes that change the MSC behavior regarding the received request such as the request number of trials, and forced attribute.

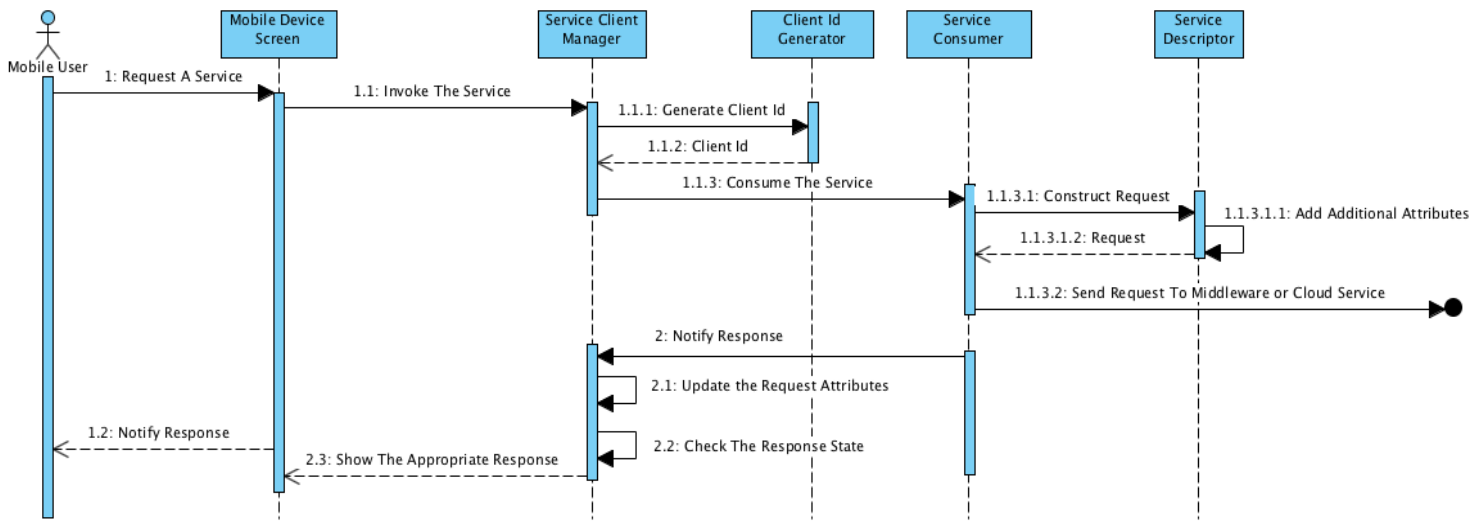

Figure 3: Cloud Service Consumer Sequence Diagram

The sequence of the client service consumer is illustrated in Figure 3. The following algorithm highlights its steps:

1- [Mobile User] select a specific service in the application to invoke.

2- [Client Id generator] generate valid client id as designed in the middleware validation method.

3- [Service Descriptor] construct the request with its specific attributes.

4- [Service Consumer] send the request to the middleware or direct to the specified cloud service as attributed.

5- [Service Client Manager] update the request attributes in the persistence storage based on the response state.

6- [Service Client Manager] adapt the response based on its state.

7- [Service Client Manager] notify the user with the appropriate response that shows the suitable state.

This sequence is performed in the mobile client part in two stages:

1- **Preparation**: The CSC prepares and constructs the appropriated request with its attributes, in order to be fit in the middleware request validation agreement.

2- **Consumption**: The stage of consuming the specific web service using the constructed request through middleware, and receiving the appropriate response to show it to the user in its reliable form.

### IV.2 Middleware Service Component (MSC)

The cloud middleware component is the other integrated part with the cloud service consumer, such that it is responsible for handling the received request and proceeding to the appropriate function to send the response back to the client consumer.

The middleware component sequence is shown in Figure 4. It begins acting once a client request is received.

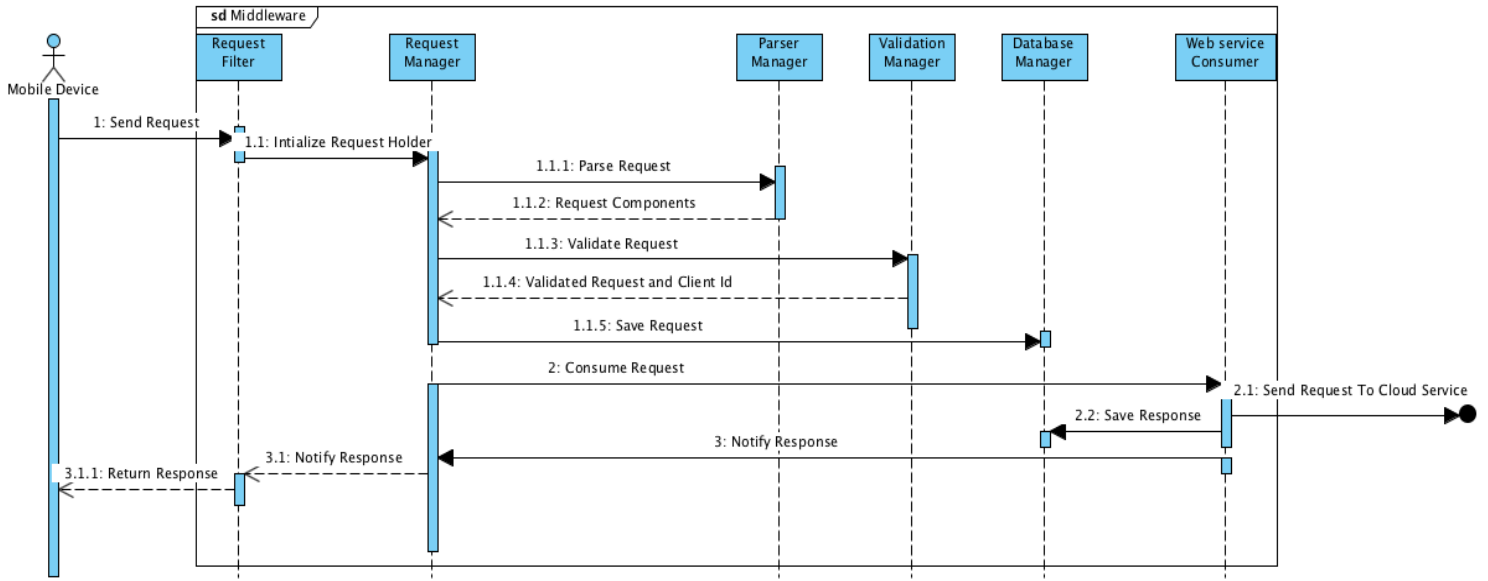

Figure 4: Service Middleware Component Sequence Diagram

The sequence of the service middleware component is illustrated in Figure 4. The following algorithm highlights its steps:

1- [Mobile device] sends a request.

2- [Middleware request filter] receive the request to ensure it is included in the allowed ones.

3- [Request manager] initialize the request holder.

4- [Parser Manager] parse the request to extract the attributes included in the request.

5- [Validation Manager] ensure the request and the attached client id attribute validity.

6- [Database Manager] save the validated request in the persistence storage.

7- [Web service Consumer] consume the attached cloud service using the sent request.

8- [Database Manager] save the response in the persistence storage with a relation with the saved request.

9- [Request Filer] respond to the mobile device with the response.

This process follows three stages,

1. **Pre-Processing:** The service middleware component parses the request to extract the client identifier and its attributes, then validate the request to ensure whether it is correct and secure or not.

2. **Processing:** In case of request validation success, the service middleware retrieves the similar previous cached request or save this request details in its cache if it did not exist before, then there are two possible cases to follow regarding the different request cases:

   a. Forward the request to the cloud service if:
      i. It was the first time invoking this request,
      ii. It is intended from the client consumer to retry this request, or
      iii. The request is labeled with the forced attribute.
   
   b. Return the middleware component cached results, if this request is invoked before with a successful result.

3. **Post-Processing**: The service middleware component constructs the appropriate response and caches its details according to the action taken in the processing step. It then sends this response to the mobile consumer to notify the user with the appropriate request state.

## V. RSAM Protocol and Reliability Analysis

The Enhanced Architecture achieves cloud service reliability while considering the most affected cases illustrated in Table 1.

The architecture process flow shown in Figure 5, covers these possible cases, such that it mainly depends on the middleware components in Client Consumer and Middleware to track the state of each request to be able to notify the user with the appropriate response.

The possible and effected response states shown in the following algorithm and the process flow in Figure 5, contains the following,

- Succeeded and failed states that explicitly indicate the success and failure of the request execution.
- Cached state that informs the user that the response was cached in the middleware from previous execution for the same request. This state covers two cases:
    o Preventing the duplicated request execution.
    o Optimizing the time of the request execution if it will get the same response from the cloud service, because it no longer needs to be passed to the cloud service again with its complex query.
- Doubt state counted as the most significant one that marks the request as doubt request to inform the user to contact the responsible one in order to ensure the request state to prevent its significant duplication.

Table 2: The covered states summary of Enhanced Middleware Architecture

| Mobile State | Middleware State | Cloud Service State |
|---|---|---|
| Reachable | Reachable | Reachable |
| Reachable | Reachable | Non-Reachable |
| Reachable | Reachable | Server Error |
| Reachable | Non-Reachable/ Server Error | Any |
| Non-Reachable | Any | Any |
| Timed Out | Reachable | Reachable |

Table 2 contains the following states:

- **Reachable**: Available through Internet access.
- **Server Error**: The Service responds with an internal server error.
- **Timed Out**: The Service execution takes more time than that allowed in the consumer to get the response.

```
1   // Cloud Service Consumer (CSC)
2   consume(request) {
3   
4       redirectToMiddleware()
5       if(networkIsReachable) { // Client Consumer
6   
7           if(middlewareIsReachable) { // Middleware
8   
9               if(middlewareError) { // Middleware
10  
11                  saveInLocalDatabase(request)
12              } else {
13  
14                  proceedInMiddleware(request)
15                  response = waitForResponse()
16                  displayResponse(response)
17              }
18          } else {
19              saveInLocalDatabase(request)
20          }
21      } else {
22          saveInLocalDatabase(request)
23      }
24  }
25  
26  // Middleware Service Component (MSC)
27  proceedInMiddleware(request) {
28  
29      if(isFirstTime(request)) { // Middleware
30  
31          saveInMiddlewareDatabase(request)
32          if(cloudServiceIsReachable) { // Cloud Service
33              proceedWithRequest(request)
34          } else {
35              failedResponse = prepareResponse(FAILED_STATE)
36              notifyClientWithReponse(failedResponse)
37          }
38      } else {
39  
40          storedRequest = getRequestFromMiddlewareDatabase(request)
41          if(isValideRequest(storedRequest)) {
42              if(hasSuccessResponse(storedRequest)) {
43                  succeededResponse = getLatestSucceededResponse(storedRequest)
44                  cachedResponse = prepareResponse(CACHED_STATE, succeededResponse)
45                  notifyClientWithReponse(cachedResponse)
46              } else {
47                  proceedWithRequest(storedRequest)
48              }
49          } else {
50              doubtResponse = prepareResponse(DOUBT_STATE)
51              notifyClientWithReponse(doubtResponse)
52          }
53      }
54  }
55  
56  proceedWithRequest(request) {
57  
58      proceedInCloudService(storedRequest)
59      response = waitForResponse()
60      saveInMiddlewareDatabase(response)
61      if(isErrorResponse(response)) {
62          failedResponse = prepareResponse(FAILED_STATE)
63          notifyClientWithReponse(failedResponse)
64      } else {
65          succeededResponse = prepareResponse(SUCCESS_STATE, response)
66          notifyClientWithReponse(succeededResponse)
67      }
68  }
69  
70  // Cloud Services
71  proceedInCloudService(request) {
72  
73      response = executeRequest(request)
74      notifyMiddlewareWithReponse(response)
75  }
```

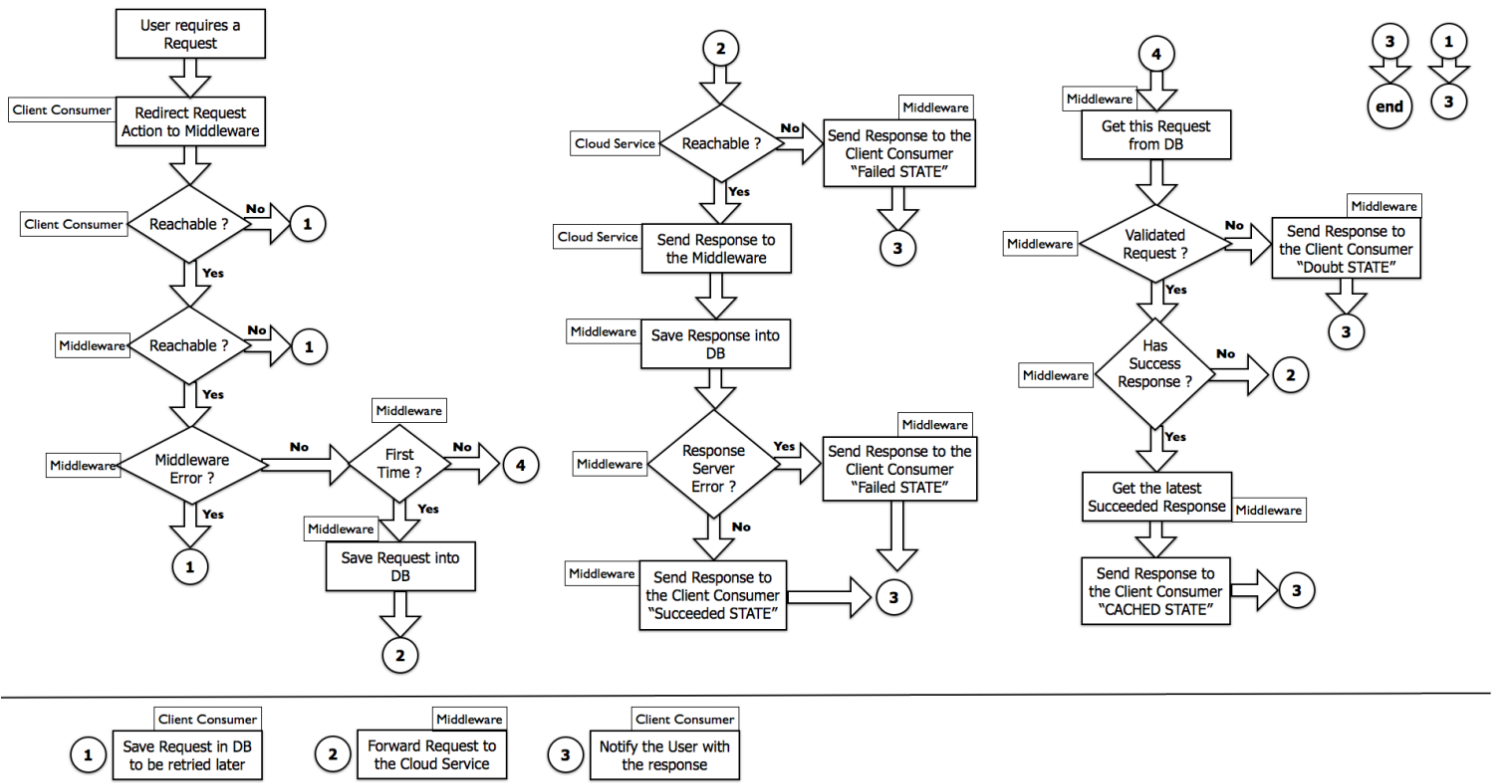

Figure 5: Enhanced Architecture Process Flow

## VI. Environment Setup

The mentioned middleware architecture is applied in the following proof of concept case, such that a social network mobile application called "Social Contacts" allows the user to send post message to his/her mobile contacts, and show customized posts regularly during whole day. This application includes the combination of the different request types and different amount of data flow as shown in Table 4, and will be clarified below.

The Social Contacts modules included in this use case are as shows in Figure 6:

1. **Cloud Services Modules**

    - **Authentication Module:** Allows the user to login to an existing account or register a new account.

    - **Feeds Module:** Shows the user the frequently posts feeds from his contacts, and allow the user to send a new post to the contacts.

2. **Middleware Module:**

    - **Request States Module:** This is a middleware related module that gains the benefits from the enhanced architecture and introduces a new mobile user experience to the sort of reliable applications, such that it contains all sent requests states to presented and be clear to the user action.

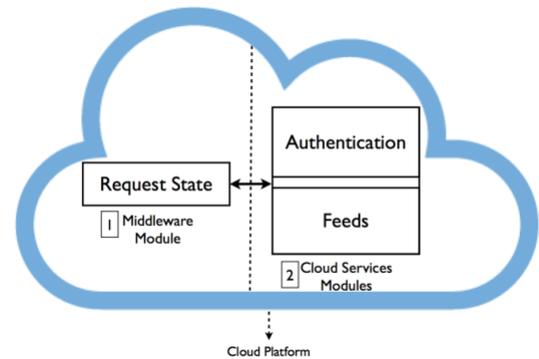

Figure 6: Social Contacts Modules

The Social Contacts application is a real application that represents the most common behavior of the functionality used in mobile applications nowadays. It includes GET, POST, and DELETE request types with variation in the request and the response data sizes ranging from small to large transferred date, through the mobile network bandwidth.

The working environment of the programming languages and the tools that used in building this enhanced architecture and the Social Contacts application functions explained as follows in Table 3,

Table 3: Environment Setup

| Cloud Services and Middleware Services Component (MSC) | | | |
|---|---|---|---|
| **Language & Frameworks** | **Web service** | **Database** | **Text Format** |
| Java Enterprice Edition & Hibernate Framework [14] | RESTFull (JAX-RS) | MySQL (V.5.2) [15] | JSON [16] |
| **Social Contacts application and the Cloud Service Consumer (CSC)** | | | |
| **Platform** | **Response Caching** | | **Text Format** |
| Android [17] | File System | | JSON |
| **Service Deployment** | | | |
| **Cloud Service Provider** | | **Application Server** | |
| Openshift Cloud provider [18] for both Middleware and Backend each on separate instance | | Glassfish Application server (V.3.1.2) [19] | |

**Note**: The CSC is built as an independent and separate android library.

## VII. Results and Analysis

The direct connection to the cloud service is the abstract connection that has no overhead neither in the transferred data nor in the connection layers. This direct connection is the one that is used through mobile clients to consume a required web service, which it is the optimal one regarding the request/ response data size and consuming time factors.

The proposed middleware approach rather than the direct connection to the cloud service, it has the RSAM protocol that discussed in Section V with its algorithm to show how it ensures the consumption reliability. It handles the abnormal flows that threat the reliable concept to be achieved though the consumption process. The measurements are considered as a comparison between the direct connection to the cloud service versus the connection through the proposed middleware components. This performance comparison will be considered to ensure the usability of the RSAM proposed approach and discuss the trade off factors with the direct cloud service. The following factors are considered as the most effective factors regarding to the mobile computing scale:

- Request size: The size of the request body that construct and send in the mobile client, it will be affected because of the additional attributes necessarily added for the middleware to ensure the reliable request.

- Response size: The size of the response body that it is received in the mobile client, it won't vary because there are no additional properties need to be added from the middleware for any reason.

- Consuming time: The time taken from sending the request till receiving its response may be longer because of the additional middleware connection rather than the direct cloud service connection.

Regarding the social contacts application functions, the following table 5, figures 7, 8, 9, show the performance measurements through middleware component and through direct cloud service connection.

The measurements indicate the middleware component cost performance. Regarding the request size factor, the middleware adds additional 226 bytes to the request. These bytes used for the required attributes discussed in Section IV. Although this number of additional bytes in the request is considered a cost to the mobile client, but this cost is low and its complexity is $O(1)$ as it is constant regardless the original request size.

The response size is the factor that is not changed, as shown in Table 5, Response Size columns. The middleware forwards the response exactly as returned from the cloud service. Therefore, we avoid using transformation overhead and additional cost from the middleware or the mobile client.

The following equations show the consuming time factor regarding both architectures: the middleware component and using the direct cloud service connection.

$$TM = Tmm + Tmc$$

$$TC = Tmc$$

Where, **TM** is the time required to consume service through middleware.

**TC** is the time required to consume service through the direct cloud service.

**Tmm** : The consumed time for the connection between the mobile client and the middleware.

**Tmc** : The consumed time for the connection between the middleware and the cloud computing.

The consuming time is one of the most important factors for the mobile applications, that requires quick responses. The measurements show that time may slightly vary about 1-2 seconds in the medium response size (~2MB) data class and the very large response (~7MB) data class. However it is exactly the same in the other cases without any overhead due to the middleware connection. This is because the middleware uses lightweight connections and lightweight data formats. In addition to its concerns with the request and response states resolving, the middleware does the minimal efforts in data transformation and data conversion.

Table 4: Social Contacts Application Functions

| Module | Function Name | Flow | Request Type | Data Size | Forced Attribute |
|---|---|---|---|---|---|
| **Authentication Module** | **Login** | • The user enters his phone number and chosen password.<br>• If the user authorized, The cloud service gets the saved user account to the mobile response, otherwise an error response is sent to the mobile client. | GET | Small (Only Phone number and Password) | YES |
| | **Register** | • The user enters his phone number and chosen password.<br>• The application takes the permission to read the contacts info and sends them to the server.<br>• If the user aren't registered before, The cloud service posts and inserts the user account with its related contacts into database, otherwise an error response is sent to the mobile client. | POST | Medium (Phone number, Password, and all contacts numbers) | NO |
| **Feeds Module** | **Send Post** | • The user enters/chooses his/her contact phone number to send the post to.<br>• The user writes his post text.<br>• The cloud service add this post in the database related with the sent contact. | POST | Small- Large (Regarding the post content size) | NO |
| | **Delete Post** | • The user chooses to delete any post appears in his/her posts.<br>• The cloud service delete this post from the database. | DELETE | Small (The post id) | NO |
| | **Get Posts Feed** | • The user opens or refresh the screen to show the posts that sent from and to him.<br>• The cloud service gets all the related posts to this user. | GET | Medium-Large (Regarding the number of returned posts) | NO |
| **[Middleware Component] - Request States Module** | **Show Requests** | • The user opens or refreshes the screen to show the requests states and the request details sent from him using any of his/her mobile devices.<br>• The middleware service gets all the related requests with their exact states to this user. | | | |
| | **Retry Request** | • The user chooses to retry the execution of specific failed request.<br>• The middleware service posts the request in its reliable corresponding scenario. | | | |
| | **Delete Request** | • The user chooses to delete any of his/her sent requests.<br>• The middleware service deletes this request from the database, in order not to be shown or retried from the user another time. | | | |

Table 5: Social Contacts Application Performance Measurements

| Module | Function Name | | Direct Cloud Connection | | | Middleware Connection | | |
|---|---|---|---|---|---|---|---|---|
| | | | Request Size (Byte) | Response Size (Byte) | Consuming Time (Second) | Request Size (Byte) | Response Size (Byte) | Consuming Time (Second) |
| Authentication Module | Login | Success | 55 | 5 | > 0 | 281 | 5 | > 0 |
| | | Failure | 53 | 315 | > 0 | 279 | 315 | > 0 |
| Feeds Module | Send Post | | 20217 (~20K) | 3072 (~3 KB) | 1 | 20443 (~20.5K) | 3072 (~3KB) | 1 |
| | Get Posts Feed | | 25 | 191745 (~ 191KB) | 2 | 251 | 191745 (~191KB) | 2 |
| | | | | 2167000 (~2MB) | 21 - 26 | | 2167000 (~2MB) | 22- 28 |
| | | | | 4592949 (~4.5MB) | 44 | | 4592949 (~4.5MB) | 44 |
| | | | | 6828571 (~7MB) | 65 – 66 | | 6828571 (~7MB) | 65 – 67 |

Direct cloud service results **time out in each time** in the large response size with heavy required computations, the middleware connection results **time out for the first time**, but return the response successfully from the middleware storage in the next retry.

This test occurred under conditions of wireless connection with 109 -115 KB/S rate, and assume the request timeout is 45 Seconds.

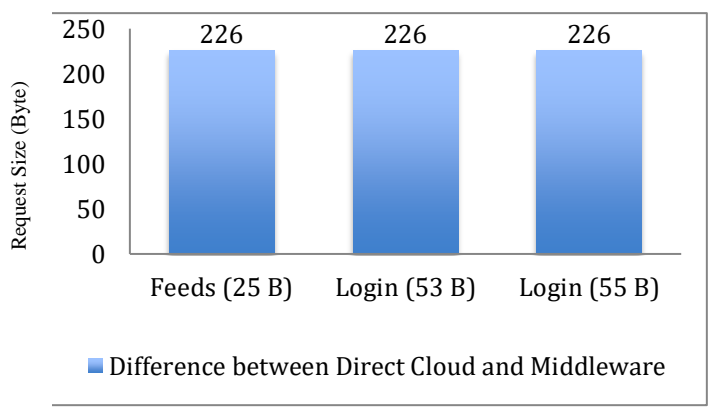

Figure 7: Extra Request Size factor of different services comparing the direct cloud with the middleware connection

Figure 7, The X-axis shows the experimental tries of different web services that varies in the request data size as: (25, 53, 55 Bytes), and the extra request data size that needed from Middleware component is shown on the Y-axis as: (226 extra Bytes for all services).

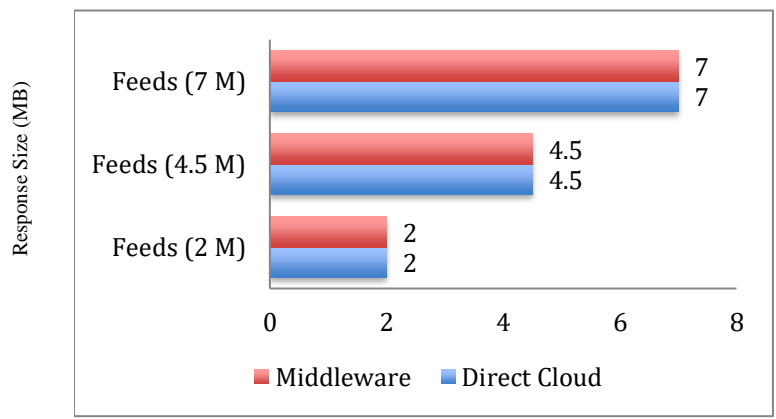

Figure 8: Response Size of different services comparing the direct cloud with the middleware connection

Figure 8, The X-axis shows the difference in the response size between Middleware component versus direct cloud, Y-axis shows the experimental tries of different web services that varies in the response data size as (2 MB, 4.5 MB, 7 MB) as shown above from Table 5.

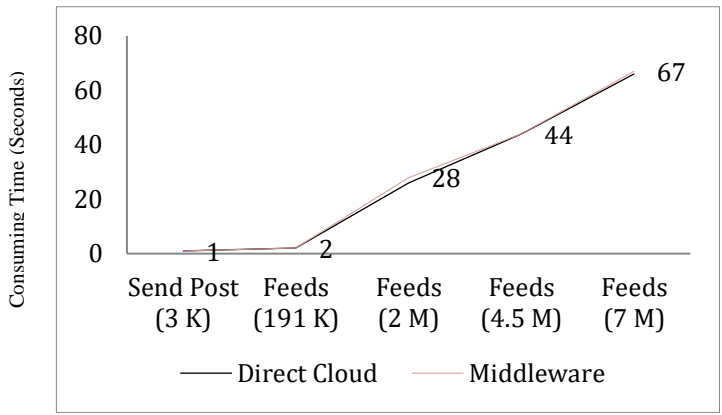

Figure 9: Consuming Time of different services comparing the direct cloud with the middleware connection

Figure 9, The X-axis shows the experimental tries of different web services that varies in the response data size as (3 KB, 191 KB, 2 MB, 4.5 MB, 7 MB) as shown in Table 5 above, and the consuming time of Middleware component versus the direct cloud is shown on the Y-axis.

## VIII. Conclusion and Future work

The middleware architecture achieves the web service consumption reliability in different cases and different environments. The proposed architecture Reliable Service Architecture using Middleware (RSAM) achieves the reliability by focusing on the request behavior rather than the request structure. In addition, it considers the most important factors for the mobile client such as the request size and the response size for mobile client data transmission limitations, and the service consuming time, which is critical for the mobile applications and their usability.

The cloud services consumption by the RSAM architecture and its protocol guarantees the reliable service communication between the mobile client and the cloud service, while it avoids adding significant communication overhead compared with the traditional direct cloud service consumption with regard to the mentioned three factors.

The services response size, database relations' communication time, and the required computations in the services are continuously increasing corresponding to the application usage, which causes timeout during the consumption of these services. The RSAM solves this issue such that the timeout is occurred in the first consumption only, but it retrieves the ready response stored in the middleware storage directly in the next retry.

Future work will be invested in implementing an enhanced approach that solves the timeout issue occurs in the heavy computation services without the need to retry to get the response from the middleware. The proposed approach mainly concerns with reducing the consumption time overhead from the mobile side, and keeping the heavy services smoothly consumed without many failures. Their tests will be conducted to measure the performance in terms of the time out occurrence possibilities in the consumption of heavy services that expected to exceed the mobile allowed time out.